\documentclass[aps,twocolumn,showpacs,groupedaddress]{revtex4}
\usepackage{amssymb,amsmath}
\usepackage{graphicx,array}
\usepackage{dcolumn}
\usepackage{multirow}
\usepackage{bm}
\usepackage{subfigure}

\begin{document}
\title{The high-pressure high-temperature phase diagram of calcium fluoride \\ 
                 from classical atomistic simulations }

\author{Claudio Cazorla}
\email{ccazorla@icmab.es}
\thanks{Corresponding Author}
\affiliation{Institut de Ci\`encia de Materials de Barcelona (ICMAB-CSIC),
             Campus UAB, 08193 Bellaterra, Spain} 

\author{Daniel Errandonea}
\affiliation{Departamento de F\'isica Aplicada (ICMUV),
             Universitat de Valencia, 46100 Burjassot, Spain}

\begin{abstract}
We study the phase diagram of calcium fluoride (CaF$_{2}$) under pressure
using classical molecular dynamic simulations performed with a simple
pairwise interatomic potential of the Born-Mayer-Huggings form.
Our results obtained under conditions $0 \le P \lesssim 20$~GPa 
and $0 \le T \lesssim 4000$~K reveal a rich variety of multi-phase 
boundaries involving different crystal, superionic and liquid 
phases, for all which we provide an accurate parametrization.    
Interestingly, we predict the existence of three \emph{special} triple 
points (i.e. solid-solid-superionic, solid-superionic-superionic 
and superionic-superionic-liquid coexisting states) within a narrow and 
experimentally accessible thermodynamic range of $6 \le P \le 8$~GPa and 
$1500 \le T \le 2750$~K.
Also, we examine the role of short-ranged repulsive (SR) and long-ranged
attractive (LA) atomic interactions in the prediction of melting lines
with the finding that SR Ca-F and LA F-F contributions are most decisive.
\end{abstract}

\maketitle

\begin{figure}
\centerline{
\includegraphics[width=0.90\linewidth]{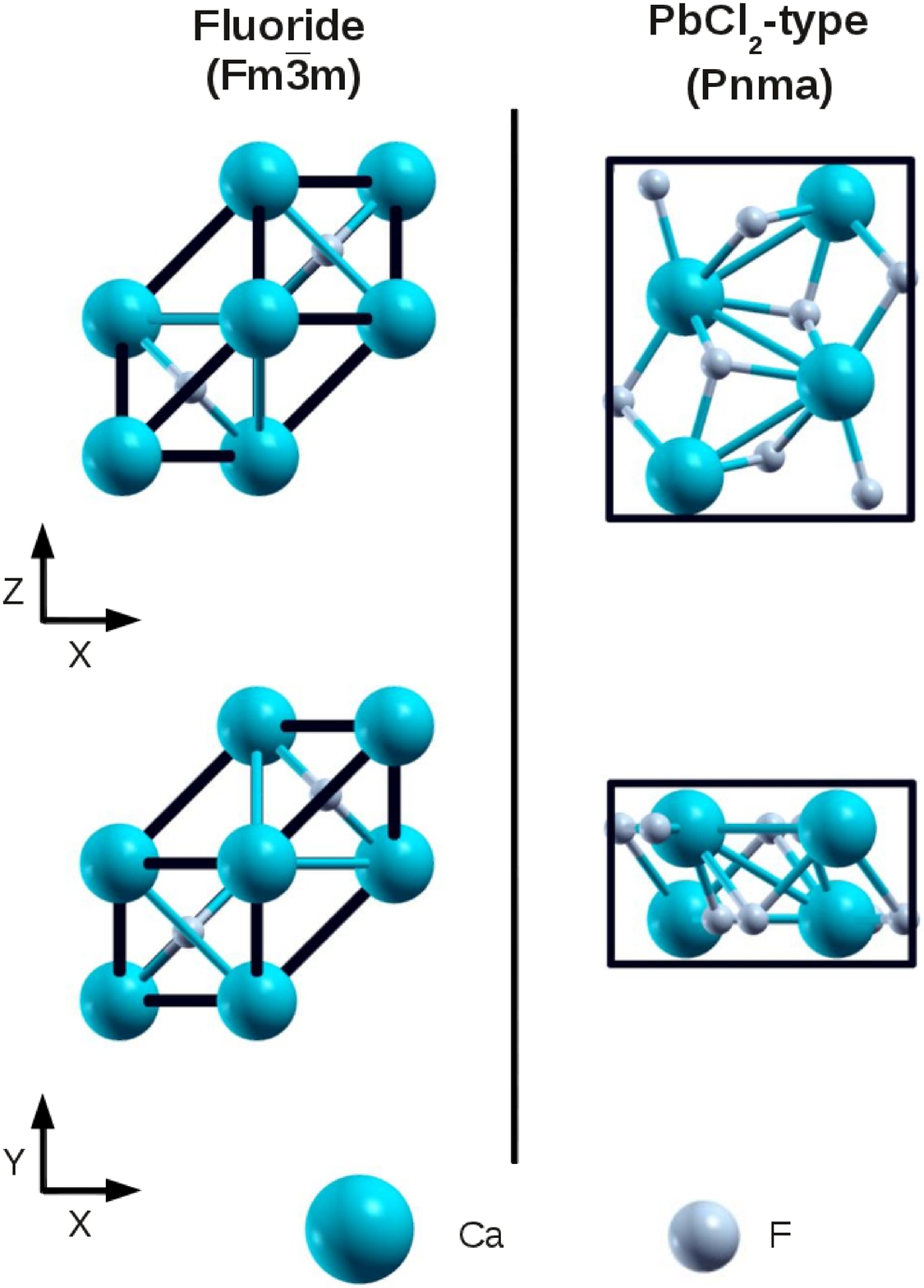}}
\caption{Cubic fluoride and orthorhombic PbCl$_{2}$-type~(contunnite) structures 
of CaF$_{2}$. In the fluoride structure Ca$^{2+}$ ions are in a cubic close-packed arrangement
and F$^{-}$ ions occupy all the tetrahedral interstitial sites.
In the contunnite structure the array of calcium cations is hexagonal close-packed (hcp)
and half the fluor anions is placed off-center in the ideally octahedral hcp voids  
with fivefold coordination (the other half of fluoride ions exhibiting tetrahedral
coordination). In the fluoride structure Ca atoms are eightfold coordinated whereas
in the contunnite structure the coordination number increases to nine. Calcium and fluor
ions are represented with blue and grey spheres respectively, and the unit cell with 
black solid lines.}
\label{fig1}
\end{figure}

\section{Introduction}
\label{sec:intro}

Calcium fluoride (CaF$_{2}$) is representative of the fluoride-structured 
halides, an important family of ionic materials with numerous applications in  
high-pressure science and technology. Examples of its assorted range of  
qualities include, excellent optical transmission properties over a wide wavelength 
range, a large electronic band gap, and very high elastic 
compressibility~[\onlinecite{barth90,gan92,verstraete03,shi05,sata02}].       
CaF$_{2}$ is also well known for being a fast-ion conductor, a material in which the  
lighter ions (i.e. F$^{-}$) acquire a significant mobility comparable to ionic melts 
at temperatures well below its fusion point~[\onlinecite{hayes85,gillan90,lindan93}].
This unusually large diffusion of fluor anions through the almost rigid 
matrix of calcium cations (i.e. Ca$^{+2}$), an effect also termed as superionicity, 
is originated by the formation of Frenkel pairs and migration of interstitial 
ions~[\onlinecite{gillan90,lindan93,wilson96,montani94}]. Superionicity, which may be 
also found in other oxide- , hydride- and iodide-based complexes like Y$_{2}$O$_{3}$,
LiBH$_{4}$ and AgI~[\onlinecite{lehovec53,matsuo07,shevlin12,stuhrmann02}], find an use 
in solid state applications as supercapacitors, batteries and fuel cells.     

The intrinsic variety of condensed matter phases in CaF$_{2}$ (i.e. solid, \emph{superionic} 
and liquid) \emph{a priori} suggests a rich and very intriguing $P-T$ phase diagram. 
Nevertheless, most of the experimental and theoretical investigations performed so far have 
focused on very narrow low-pressure and low-temperature thermodynamic ranges. 
The $T=0$ polymorphism of CaF$_{2}$ encompasses two main crystal structures (see Fig.~\ref{fig1}), 
the low-pressure face-centered cubic fluoride phase (space group $Fm\bar{3}m$) and 
the high-pressure orthorhombic PbCl$_{2}$-type (cotunnite) phase (space group
$Pnma$). Although CaF$_{2}$ fast-ion conduction has been intensively investigated  
at ambient pressures~[\onlinecite{hayes85,gillan90,lindan93,wilson96,montani94}], 
it surprisingly remains yet unexplored in the cubic and orthorhombic phases under compression.
Gathering information on the pressure dependence of CaF$_{2}$ fast-ion conduction 
and melting however turns out to be very desirable both from a fundamental and technological 
point of view. Assuming fast-ion conduction behavior in both low- and high-pressure phases, 
for instance, opens the possibility for the existence of \emph{special} triple or even quadruple 
points (i.e. thermodynamic states in which several superionic and solid phases might coexist 
in thermodynamic equilibrium) at elevated $P-T$ conditions. Also, our present knowledge 
on the fusion properties of CaF$_{2}$ and of ionic compounds in general, is unfortunately 
rather scarce~[\onlinecite{zijiang05,wang10,zeng08}].
Likewise, the expected connections between atomic sub-lattice melting and homogeneous melting, 
a subject that we have recently investigated in Ar(H$_{2}$)$_{2}$~[\onlinecite{cazorla10}], are not 
yet clearly understood. On the practical side, pressure-induced trends unravelled in 
CaF$_{2}$ could be generalized to other fluoride-structured materials composed of heavy 
and light ion species, that have been proposed or are used in relevant technological 
applications (e.g. transition metal hydrides in hydrogen storage 
cells and uranium dioxide for generation of nuclear fuel).     

In this article, we report the phase diagram of CaF$_{2}$ under pressure 
(e.g. $0 \le P \lesssim 20$~GPa) as obtained from classical molecular dynamics (MD) 
simulations performed with a simple but reliable rigid-ion interatomic potential of 
the Born-Mayer-Huggings (BMH) form. Our results rely on extensive one- and two-phase 
coexistence simulations and reveal a rich variety of previously unreported $P-T$ phase 
boundaries~[\onlinecite{zijiang05,wang10,zeng08}] 
(i.e. fluoride-PbCl$_{2}$, fluoride-liquid, PbCl$_{2}$-liquid,
fluoride-superionic fluoride, PbCl$_{2}$-superionic PbCl$_{2}$, superionic fluoride-PbCl$_{2}$ 
and superionic fluoride-superionic PbCl$_{2}$), for all which we provide here an accurate 
parametrization. For example, we find that both fluoride-superionic fluoride 
and PbCl$_{2}$-superionic PbCl$_{2}$ phase boundaries, $T_{s} (P)$, are linearly 
dependent on pressure with a small and positive $dT_{s}/dP$ slope of $34.2$ and $50.2$~K/GPa, 
respectively. Interestingly, we predict the existence of three \emph{special} triple points 
involving coexistence of fluoride-superionic fluoride-PbCl$_{2}$, superionic 
fluoride-PbCl$_{2}$-superionic PbCl$_{2}$ and superionic fluoride-superionic PbCl$_{2}$-liquid 
phases, within a narrow and experimentally accessible region of 
$6 \le P \le 8$~GPa and $1500 \le T \le 2750$~K.  
In addition to these findings, we use a free-energy perturbative approach to 
evaluate the shift in melting temperature caused by mild variations of the employed 
potential parameters. By doing this, we quantify the role of repulsive and dispersion
interactions on our results and qualitatively gain access to the melting features of 
BMH potentials used in other works.

The organization of this article is as follows. In the next section, we describe the 
computational methods employed and the low-temperature performance of the selected 
Born-Mayer-Huggings potential. In Sec.~\ref{sec:results}, we present our results 
for the phase diagram of CaF$_{2}$ under pressure and discuss them. 
Then, a section follows in which we analyze the role of repulsive and dispersion 
interactions on the determination of melting points by theoretical means. 
Finally, we summarize the main conclusions in Sec.~\ref{sec:conclusions}.

\section{Simulation Details and The Interatomic Pair Potential}
\label{sec:simulation} 

Calculations were done with LAMMPS~[\onlinecite{lammps}], a parallel classical molecular 
dynamics (MD) code comprising a large variety of potentials and different schemes 
for simulation of solid-state and soft materials. 
Our MD calculations are of two main types: one-phase (i.e. pure liquid, solid
and superionic phases) and two-phase coexistence (i.e. liquid and superionic phases coexisting in 
thermodynamic equilibrium) simulations. 
One-phase simulations are performed in the canonical $( N, V, T )$ 
ensemble and two-phase coexistence simulations in the microcanonical $( N, V, E )$ ensemble 
(specific details of these simulations are provided in Sec.~\ref{sec:results}). 
In $( N, V, T )$ simulations the temperature is kept fluctuating around a constrained 
value by using Nose-Hoover thermostats. Large simulation boxes containing $6,144$ 
and $12,288$ atoms are used in our one-phase and two-phase coexistence simulations, 
respectively. Periodic boundary conditions are applied along the three Cartesian directions 
in all the calculations. Newton's equations of motion are integrated using the customary 
Verlet's algorithm and a time-step length of $10^{-3}$~ps. A particle-particle particle-mesh 
$k$-space solver is used to compute long-range van der Waals and Coulomb interactions beyond 
a cut-off distance of $12$~\AA~ at each time step. 

\begin{equation}
V_{ij}(r) = A_{ij}e^{-\frac{r}{\rho_{ij}}} - \frac{C_{ij}}{r^{6}} + \frac{Z_{i}Z_{j}}{r}  
\label{eq:BMH}
\end{equation}

\begin{table}
\begin{center}
\label{tab:potential} 
\begin{tabular}{ c c c c }
\hline
\hline
$  $ & $  $ & $  $ & $  $ \\
$  $ & $\quad A~{\rm (eV)} \quad $ & $\quad \rho~{\rm (\AA)} \quad $ & $\quad  C~{\rm (eV~\AA^{6})} $ \\
$  $ & $  $ & $  $ & $  $ \\
\hline
$  $ & $  $ & $  $ & $  $ \\
$ {\rm Ca-F}   $ & $ 1717.441 $ & $ 0.287 $ & $ 0.102  $ \\
$  $ & $  $ & $  $ & $  $ \\
$ {\rm F-F}   $ &  $ 2058.994 $ & $ 0.252 $ & $ 16.703 $ \\
$  $ & $  $ & $  $ & $  $ \\
\hline
\hline
\end{tabular}
\end{center}
\caption{Interatomic pair potential parameters for CaF$_{2}$~[\onlinecite{dick58}].}
\end{table}

The interatomic potential adopted for this study is 
$U (r) = V_{\rm CaF} (r) + V_{\rm FF} (r)$ where terms $V_{ij}$
are of the Born-Mayer-Huggings (BMH) form (see Eq.~\ref{eq:BMH}). 
Each pairwise term is composed of three different contributions; 
the first is of exponential type and accounts for the short-ranged 
atomic repulsion deriving from the overlapping between different electron clouds; 
the second term is proportional to $r^{-6}$, with $r$ being the radial distance between a 
given couple of ions, and represents the long-ranged atomic attraction due to dispersive 
van der Waals forces; the third term is the usual Coulomb interaction between puntual atomic 
charges, which in our case are taken to be $Z_{\rm Ca} = +2e$ and $Z_{\rm F} = -1e$. 
In Table~I, we enclose the value of the BMH parameters used throughout this work and 
which coincide with those primarily deduced by Dick and Overhauser~[\onlinecite{dick58}].  
It must be noted that the original Dick-Overhauser potential includes electronic 
polarization effects via a shell model however for present purposes 
these can be safely neglected due to the marked ionic nature of CaF$_{2}$. 
Indeed, Lindan and Gillan explicitly showed for a similar BMH model that 
inclusion of electronic polarizability had a remarkable small effect on the estimation 
of static and dynamic CaF$_{2}$ quantities~[\onlinecite{gillan90,lindan93}].    

\begin{figure}
\centerline{
\includegraphics[width=1.00\linewidth]{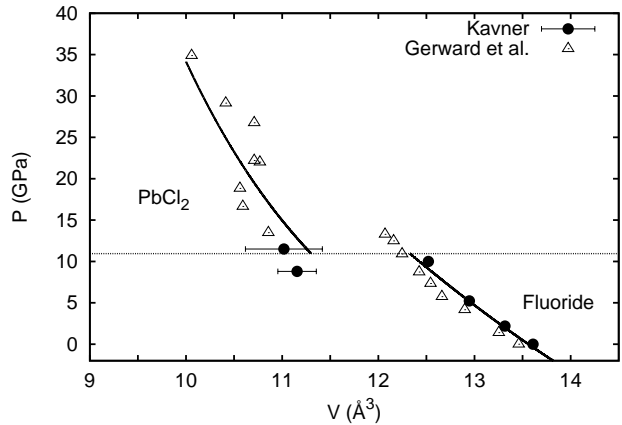}}
\caption{Zero-temperature equation of state of CaF$_{2}$ under pressure obtained
with the present BMH potential (solid lines). Experimental data can be found in 
Refs~[\onlinecite{kavner08}] and [\onlinecite{gerward92}].}
\label{fig2}
\end{figure}

In order to assess the reliability of the adopted BMH potential, we performed a series 
of static ground-state calculations and compared them with available low-$T$ experimental data. 
In Fig.~\ref{fig2}, we show the zero-temperature equation of state of CaF$_{2}$ 
obtained for its cubic and orthorhombic phases (solid lines). In each phase, we computed the 
energy per formula unit for a set of $20$ volume points spanning over the 
range $9.0 \le V \le 14.5$~\AA$^{3}$. Subsequently, we 
fitted the results to a third order Birch-Murnaghan equation of the form
\begin{eqnarray}
&&E_{\rm perf}( V ) = E_{0} + \frac{3}{2}~V_{0}~K_{0}~\cdot \nonumber \\
&& \bigg [ -\frac{\chi}{2} \left ( \frac{V_{0}}{V} \right )^2 + \frac{3}{4}~ \left ( 1+2 \chi \right ) \left ( \frac{V_{0}}{V} \right )^{(4/3)} \nonumber \\
&& - \frac{3}{2} \left ( 1+\chi \right ) \left (\frac{V_{0}}{V} \right )^{(2/3)} + \frac{1}{2} \left (\chi+\frac{3}{2}\right ) \bigg ]~,
\label{eq:eqstate}
\end{eqnarray}
where $E_{0}$ and $K_{0}=-V_{0}\frac{d^2E}{dV^2}$ are the energy and bulk modulus at 
equilibrium volume $V_{0}$, $\chi = \frac{3}{4}\left ( 4 - K^{'}_{0} \right )$ 
and ~$K_{0}^{'}=\left [\partial K/\partial P\right ]$, with derivatives taken at zero 
pressure. (Atomic forces and cell shape relaxations were performed for the 
orthorhombic PbCl$_{2}$-type phase at each volume.) The value of the 
resulting $E_{0}$, $V_{0}$, $K_{0}$ and $K_{0}^{'}$ parameters are $-9.05$~($-8.97$)~eV, 
$13.56$~($12.52$)~\AA$^{3}$, $108.1$~($79.5$)~GPa, and $1.51$~($5.81$) for the 
cubic (orthorhombic) structure. The static equation of state then is obtained 
as the minus derivative of Eq.~(\ref{eq:eqstate}). By comparing the enthalpy of 
the different phases we find that the zero-temperature cubic~$\to$~orthorhombic phase 
transition occurs at a pressure of $P_{t} = 10.95$~GPa, in very good agreement with 
recent ambient experimental data obtained by Kavner and 
others~[\onlinecite{kavner08,gerward92}]. The corresponding change of volume
is $-8.35$~\% with $V = 12.33~(11.30)$~\AA$^{3}$ in the cubic~(orthorhombic)
phase.  

In Table~II, we report the bulk modulus and structural parameters of the 
two polymorphisms of CaF$_{2}$ obtained at equilibrium and $P_{t}$. As one can see, the 
accordance with experiments in this case is also notable. It must be noted that 
from the Birch-Murnaghan fit quoted above we obtain a zero-pressure bulk modulus that is 
$\sim 20$~GPa larger than the experimental value reported by Dorfman 
\emph{et al.}~[\onlinecite{dorfman10}]. 
However, in the obtaining of that experimental datum the value of 
$K_{0}^{'}$ parameter in the corresponding equation of state fit was set to $4.7$. 
Proceeding in the same way as Dorfman and collaborators, we obtain $K_{0} = 82$~GPa 
which is in very close agreement with their measurements.   
Finally, we computed the value of the three independent elastic constants of the $Fm\bar{3}m$ phase at 
equilibrium, $C_{ij}$'s. For this, we distorted the shape of the cubic unit cell 
according to the strain matrices shown in Table~III and fitted the resulting variation of 
the energy to the also reported parabolic curves~[\onlinecite{shi09}]. 
We obtain $C_{11} = 168.1$, $C_{12} = 46.9$, and $C_{44} = 37.7$~GPa which agree notably 
with the experimental values $C_{11}^{expt} = 165.4$, $C_{12}^{expt} = 44.4$, 
and $C_{44}^{expt} = 34.2$~GPa~[\onlinecite{catti91}], and previous theoretical 
estimations as well~[\onlinecite{zeng08}].               

The main conclusion emerging from these calculations is that the adopted 
BMH potential provides a very reliable account of the low-$T$ region of the
phase diagram of CaF$_{2}$. It can then be assumed that medium and high-temperature 
descriptions obtained with this same simple model interaction will be also physically 
meaningful. In fact, as we will show in the next section, superionic and melting 
temperatures predicted at ambient pressures are fully consistent with 
experimental observations.          

\begin{table*}
\begin{center}
\label{tab:structure} 
\begin{tabular}{@{\hspace{1.0cm}}c@{\hspace{1.0cm}}c@{\hspace{1.0cm}}c@{\hspace{1.0cm}}c@{\hspace{1.0cm}}c@{\hspace{1.0cm}}} 
\hline
\hline
$ $ & $ $ & $ $ & $ $ & $ $ \\
$ $ & \multicolumn{2}{c}{${\rm Atomic~Structure}$~(\AA)} & \multicolumn{2}{c}{${\rm Bulk~Modulus~(GPa)}$} \\ 
$ $ & \multicolumn{2}{c}{$ $} & \multicolumn{2}{c}{$ $} \\ \cline{2-3} \cline{4-5}
$ $ & $ $ & $ $ & $ $ & $ $ \\ 
$ $ & $ {\rm Theory} $ & $ {\rm Experiment} $ & $ {\rm Theory} $  & $ {\rm Experiment} $  \\ 
$ $ & $ $ & $ $ & $ $ & $ $ \\
\hline
$ $ & $ $ & $ $ & $ $ & $ $ \\
$ Fm\bar{3}m~(P=0) $ & $ a_{0} = 5.460 $ & $ a_{0} = 5.463^{a} $ & $ 108~(K_{0}^{'}=1.5) $ & $ $ \\
$ $ & $ $ & $ $ & $82~(K_{0}^{'}=4.7) $ & $85^{b}~(K_{0}^{'}=4.7) $ \\
$ $ & $ $ & $ $ & $ $ & $ $ \\
$ Fm\bar{3}m~(P=P_{t}) $ & $ a_{0} = 5.289 $ & $ a_{0} = 5.313^{b} $ & $ 122 $ & $ 152(20)^{c}  $ \\
$ $ & $ $ & $ $ & $ $ & $ $ \\
$ Pnma~(P=P_{t})       $ & $ a_{0} = 5.721 $ & $ a_{0} = 5.700^{b} $ & $ 138 $ & $ 162(30)^{c} $ \\
$                            $ & $ b_{0} = 3.463 $ & $ b_{0} = 3.450^{b} $ & $ $ & $ $ \\
$                            $ & $ c_{0} = 6.846 $ & $ c_{0} = 6.800^{b} $ & $ $ & $ $ \\
$ $ & $ {\rm Ca~(0.2472, 0.25, 0.1153)} $ & $ {\rm Ca~(0.2530, 0.25, 0.1094)^{d}} $ & $ $ & $ $ \\
$ $ & $ {\rm F1~(0.8510, 0.25, 0.0732)} $ & $ {\rm F1~(0.8595, 0.25, 0.0731)^{d}} $ & $ $ & $ $ \\
$ $ & $ {\rm F2~(0.4768, 0.25, 0.8300)} $ & $ {\rm F2~(0.4780, 0.25, 0.8344)^{d}} $ & $ $ & $ $ \\
$ $ & $ $ & $ $ & $ $ & $ $ \\
\hline
\hline
\end{tabular} 
\end{center}
\caption{Structural parameters and bulk modulus $K = -V(dP/dV)_{T}$ of CaF$_{2}$ obtained 
at zero-temperature; 
theory values are obtained with the present BMH potential and experimental uncertainties are 
indicated within parentheses. $P_{t} = 10.95$~GPa is the pressure at which the 
cubic~$\to$~orthorhombic phase transformation is found to occur. 
Experimental data can be found in Refs.~[\onlinecite{wu05}]$^{a}$, [\onlinecite{dorfman10}]$^{b}$,
[\onlinecite{gerward92}]$^{c}$  and [\onlinecite{morris01}]$^{d}$.} 
\end{table*}

\begin{table*}
\begin{center}
\label{tab:elastic} 
\begin{tabular}{@{\hspace{1.0cm}}c@{\hspace{1.0cm}}c@{\hspace{1.0cm}}c@{\hspace{1.0cm}}} 
\hline
\hline
$ $ & $ $ & $ $ \\
${\rm Stress-strain~coefficient} $ & $ {\rm Strain~matrices} $ & $ {\rm Strain~energy} $ \\
$ $ & $ $ & $ $ \\
\hline
$ $ & $ $ & $ $ \\
$C_{11},~C_{12}$ & $ \left( \begin{matrix} 1 + \epsilon & 0 & 0 \\
                                     0 & 1 + \epsilon & 0 \\
                                     0 & 0 & 1 + \epsilon  \end{matrix} \right) $ & $u(\epsilon) = \frac{3}{2}\left( C_{11} + 2C_{12} \right)\epsilon^{2} $ \\
$ $ & $ $ & $ $ \\
$C_{11},~C_{12}$ & $ \left( \begin{matrix}  1 + \epsilon & 0 & 0 \\
                                     0 & 1 + \epsilon & 0 \\
                                     0 & 0 & \frac{1}{\left(1 + \epsilon \right)^{2}}  \end{matrix} \right) $ & $u(\epsilon) = 3\left( C_{11} - C_{12} \right)\epsilon^{2} $ \\
$ $ & $ $ & $ $ \\
$C_{44} $ & $ \left( \begin{matrix}  1 & \epsilon & 0 \\
                                \epsilon & 1 & 0 \\
                                     0 & 0 & \frac{1}{1 - \epsilon^{2}} \end{matrix} \right) $ & $u(\epsilon) = 2C_{44}\epsilon^{2} $ \\
$ $ & $ $ & $ $ \\
\hline
\hline
\end{tabular} 
\end{center}
\caption{Strain matrices for the three independent elastic constants $C_{11}$, $C_{12}$ and
$C_{44}$ of the cubic $Fm\bar{3}m$ structure and corresponding strain energy relations per 
unit volume at $P = 0$.}
\end{table*}

\section{Results and Discussion}
\label{sec:results}

\begin{figure*}
\centerline{
\includegraphics[width=0.90\linewidth]{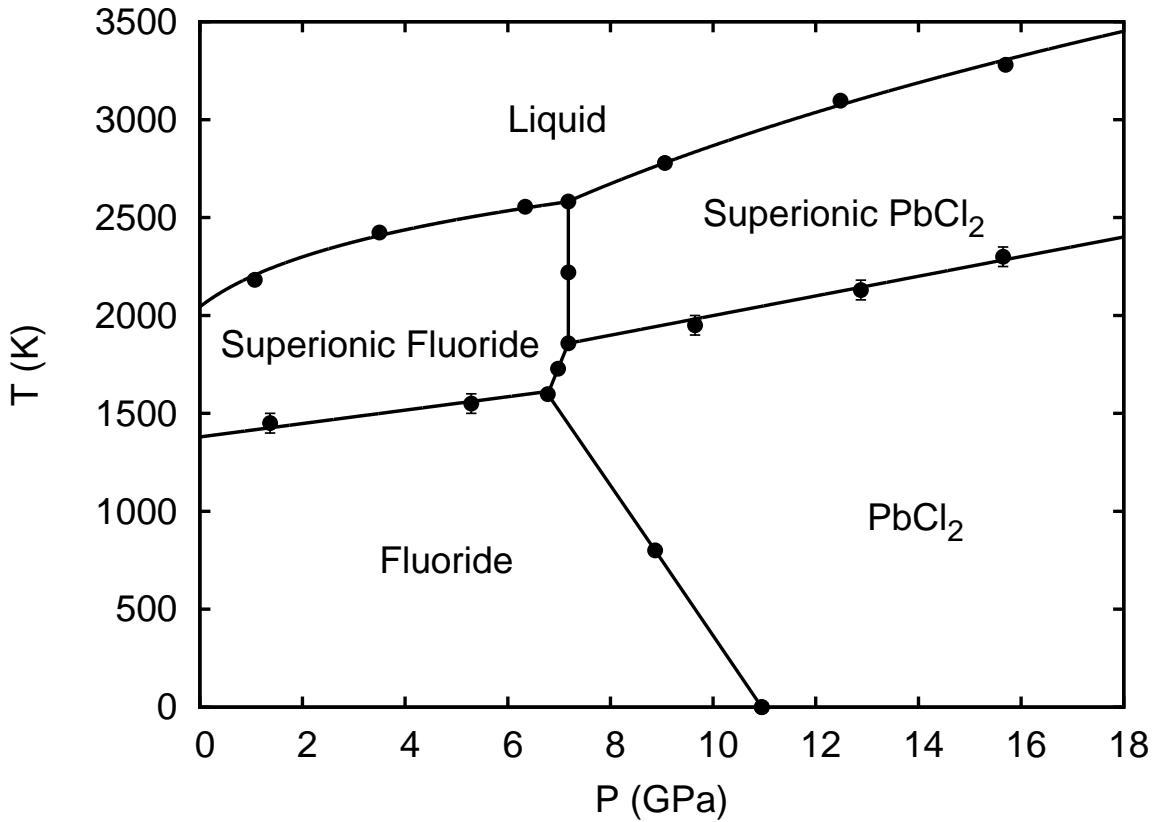}}
\caption{The phase diagram of CaF$_{2}$ under pressure obtained from molecular dynamics 
simulations. Solid lines represent function fits (see text) and filled 
dots thermodynamic states at which molecular dynamic simulations were explicitly 
carried out. Transition temperature errors are represented with solid bars.}
\label{fig3}
\end{figure*}

In the next sections we will describe in detail the two-phase boundaries
and \emph{special} triple points appearing on the phase diagram of CaF$_{2}$
shown in Fig.~\ref{fig3}, however let us first explain the general 
strategy that we followed to obtain it. 

Initially, we performed extensive one-phase and two-phase molecular dynamics 
simulations to find out the solid-superionic and superionic-liquid phase boundaries
of the fluoride and PbCl$_{2}$-type structures in the whole range of pressures 
considered (i.e $0 \le P \lesssim 20$~GPa). Next, we considered the thermodynamic state 
$(P_{1c}, T_{1c})$ at which the melting curves 
of the two superionic states cross each other. The Gibbs free energy of the superionic 
fluoride and superionic orthorhombic phases are equal at that thermodynamic state  
thus $(P_{1c}, T_{1c})$ must belong also to the boundary separating the superionic 
cubic and superionic orthorhombic regions. In other words, $(P_{1c}, T_{1c})$ is a 
\emph{special} triple point (\emph{special} because it comprises coexistence of 
superionic and liquid phases, in contrast to ordinary solid-solid-liquid and/or 
solid-liquid-vapor triple points). One-phase MD simulations were then carried out to  
determine the volume and enthalpy (i.e. $H = E + PV$) of the two superionic phases at 
$(P_{1c}, T_{1c})$. Using the customary Clausius-Clapeyron relation  
\begin{equation}
\frac{dT}{dP} = T \frac{\Delta V}{\Delta H}~, 
\label{eq:clausius}
\end{equation}
we computed the value of the superionic cubic-superionic orthorhombic
boundary slope at that triple point, and assumed it to be constant along 
the multi-phase boundary. 
(In fact, neglecting $dT / dP$ variations of this type   
may introduce some errors in our predictions however, as we will show in 
Sec.\ref{subsec:boundaries}, these turn out to be rather small.)   
By tracing the superionic fluoride-superionic PbCl$_{2}$-type phase 
boundary, a second \emph{special} triple point $(P_{2c}, T_{2c})$ is found at 
its intersection with the solid orthorhombic-superionic orthorhombic phase boundary 
(see Fig.~\ref{fig3}). Likewise, one-phase MD simulations were conducted at 
$(P_{2c}, T_{2c})$ to obtain the value of the corresponding $dT / dP$ slope 
and assumed it to be constant along the superionic cubic-solid 
orthorhombic phase boundary. 
Proceeding as before, we identified the existence of a third \emph{special} triple point 
$(P_{3c}, T_{3c})$ at the crossing of the superionic cubic-solid 
orthorhombic phase boundary with the solid fluoride-superionic fluoride phase 
boundary. Finally, the boundary separating the solid cubic and solid orthorhombic 
regions was drawn by joining the thermodynamic states $(P_{3c}, T_{3c})$ and $(P_{t}, 0)$, 
where $P_{t}$ is the pressure at which the cubic~$\to$~orthorhombic transformation 
occurs at $T = 0$. 

In what follows we describe and yield the parametrization of all the multi-phase 
boundaries cited above, explaining the simulation procedures that we followed to 
obtain them.

\subsection{Superionicity}
\label{subsec:superionicity}

Comprehensive one-phase $(N , V, T)$ molecular dynamics simulations were 
carried out to compute the solid-superionic phase boundary of cubic 
and orthorhombic CaF$_{2}$ as a function of pressure. Calculations   
comprised large simulation boxes of $6,144$ atoms and exceptionally long 
simulation times of $\sim 200$~ps. We systematically carried out simulations 
at temperature intervals of $100$~K, up to $3000$~K, at each 
volume considered.

\begin{figure}
\centerline
        {\includegraphics[width=1.0\linewidth]{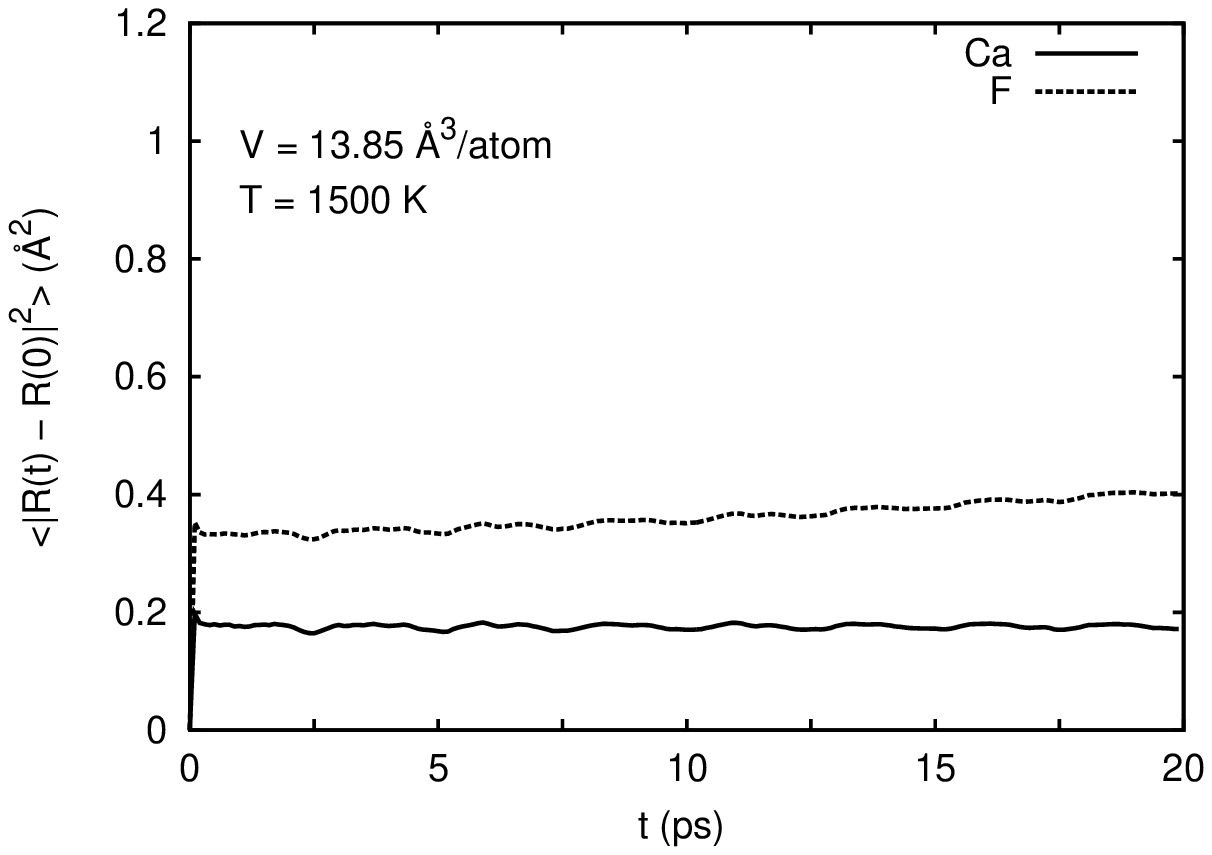}}%
        {\includegraphics[width=1.0\linewidth]{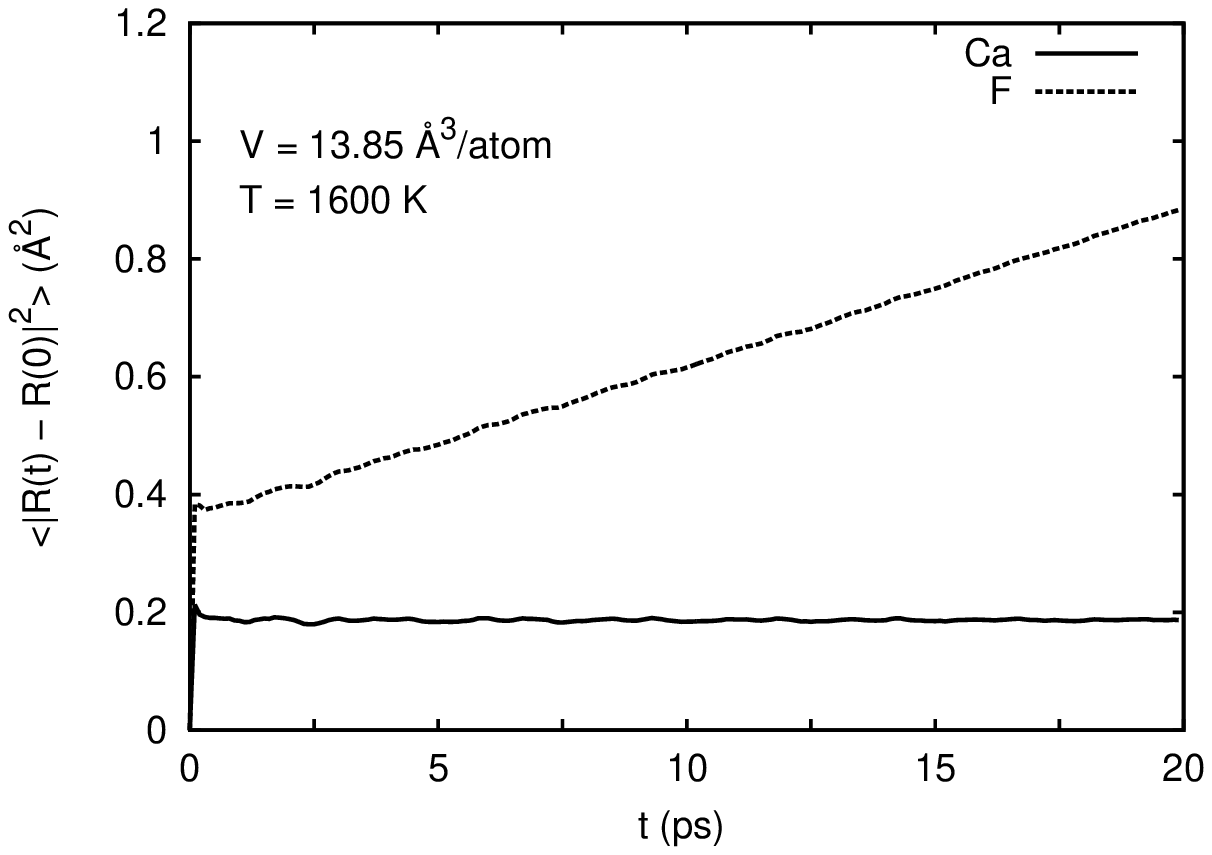}}%
        \caption{Mean squared displacement of fluor and 
                 calcium species represented as a function of time at temperatures
                 below (\emph{Top}) and above (\emph{Bottom}) the corresponding 
                 transition point $T_{s} = 1550$~K. One-phase molecular  
                 dynamics simulations were performed for cubic CaF$_{2}$ at pressures 
                 $\sim 5.3$~GPa.}
\label{fig4}
\end{figure}

\begin{figure}
\centerline
        {\includegraphics[width=1.0\linewidth]{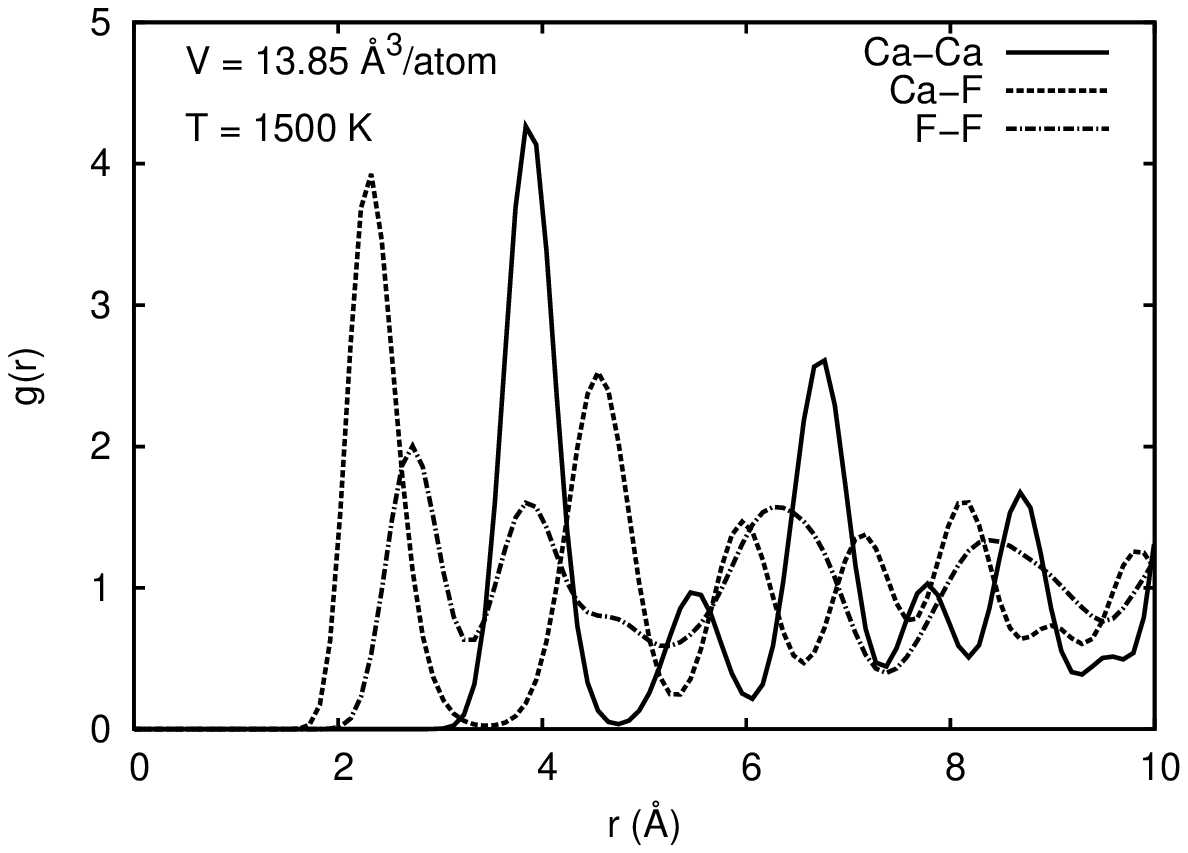}}%
        {\includegraphics[width=1.0\linewidth]{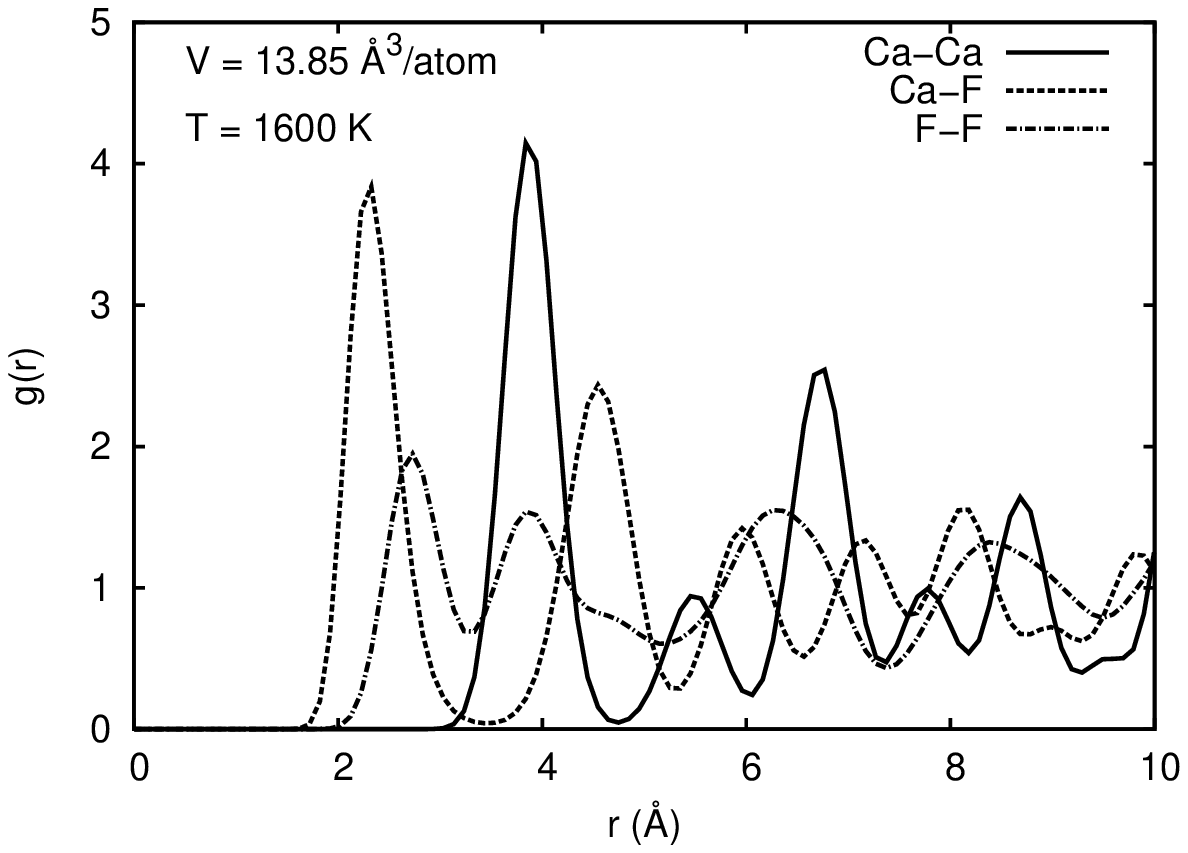}}%
        \caption{Radial pair distribution functions obtained at temperatures 
                 below (\emph{Top}) and above (\emph{Bottom}) the corresponding 
                 superionic transition point $T_{s} = 1550$~K of cubic CaF$_{2}$ 
                 at $P \sim 5.3$~GPa.}
\label{fig5}
\end{figure}

Following previous works~[\onlinecite{gillan90,lindan93,yin04,araujo09}],
we identified superionicity via inspection of the time-dependent mean 
squared displacement function (MSD) obtained in one-phase MD simulations. 
The MSD function is defined as  
\begin{equation}
\langle | \Delta R_{i}^{2}(t) | \rangle = \langle | R_{i}(t+t_{0}) - R_{i}(t_{0}) |^{2} \rangle~,  
\end{equation}
where $R_{i}(t)$ is the position of atom $i$ at time $t$, $t_{0}$ is an arbitrary time
origin, and $\langle \cdots \rangle$ denotes thermal average.  
($\langle | \Delta R_{i}^{2}(t) | \rangle$ was computed separately for each 
ionic species and thermal averages were performed over $t_{0}$ and atoms
-see Fig.~\ref{fig4}-). In practice, F$^{-}$ diffusion is signaled by the 
appearance of a non-zero MSD slope as we illustrate in Fig.~\ref{fig4}. 
It is worth noticing that anion diffusion is hardly discernable when 
scrutinizing only spatially averaged static quantities; for instance, one can not appreciate 
important differences between the CaF${_2}$ radial pair distribution functions obtained at 
temperatures just above and below the superionic transition (see Fig.~\ref{fig5}).             

For each phase, we determined the superionic transition temperature 
$T_{s} (P)$ at six different volumes. 
In both phases we found that the results could be perfectly 
fitted to straight lines $T_{s} = a_{s} + b_{s}P$, where 
$a_{s} = 1379.69$~K and $b_{s} = 34.23$~K/GPa are the optimal parameters 
for the cubic phase and $a_{s} = 1497.70$~K and $b_{s} = 50.17$~K/GPa 
for the orthorhombic phase (see Fig.~\ref{fig3}). 
As it can be observed, the slope of both fluoride and PbCl$_{2}$-type 
$T_{s} (P)$ curves are positive so indicating that the superionic phases are 
entropically stabilized over the corresponding crystals (i. e. the 
volume of the former systems is larger than those of the last -$0 < \Delta V$- 
and consequently so is the enthalpy -$0 < \Delta H$-; then $\Delta H = T \Delta S$ 
must be accomplished over the coexistence line and $0 < \Delta S$ follows). 
This result is reminiscent of customary melting however no depletion of the 
two-phase boundary induced by compression is observed in the present case.  

For ambient pressures, we obtain a superionic transition temperature of 
$T_{s}(0) = 1380~(10)$~K which is in very good agreement with  
experimental data $T_{s}^{expt} = 1420~(20)$~K found in 
Refs.~[\onlinecite{derrington75,hayes89,evangelakis89,dixon80}].
Also, we computed the diffusion coefficient of F$^{-}$ anions, $D_{-}$, from a least 
squares fit to the MSD profiles 
(i.e. $\langle | \Delta R_{i}^{2}(t) | \rangle = A_{-} + 6D_{-}t$, following the
well known Einstein relation).
Our $D_{-}$ results at temperatures $1400$, $1500$ and $1600$~K   
are $0.3$, $0.9$ and $2.7$~$10^{-5}$~cm${^2}$/s, which turn out to be consistent 
with experimental data $0.6$, $1.6$ and 
$3.2$~$10^{-5}$~cm${^2}$/s~[\onlinecite{gillan90,wilson96,derrington75,evangelakis89}].

\subsection{Melting}
\label{subsec:melting}

\begin{figure}
\centerline
        {\includegraphics[width=1.0\linewidth]{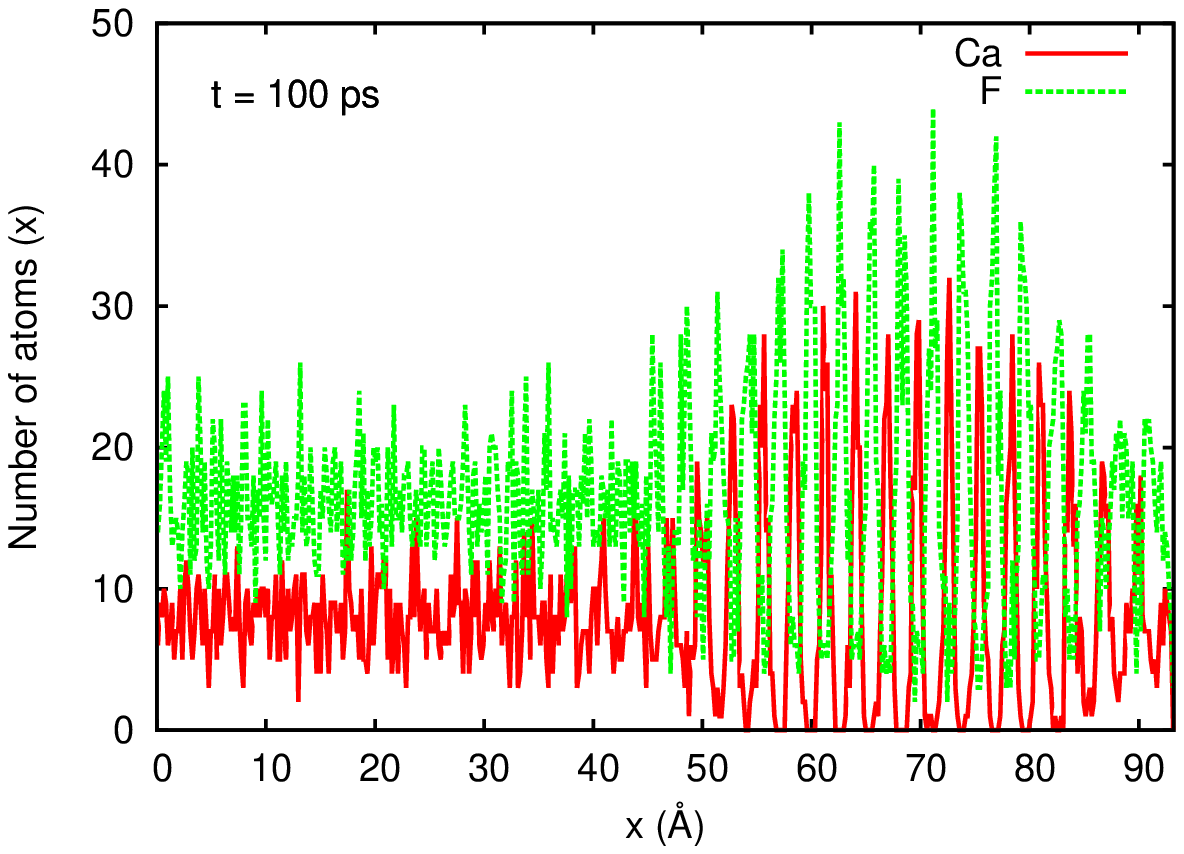}}%
        {\includegraphics[width=1.0\linewidth]{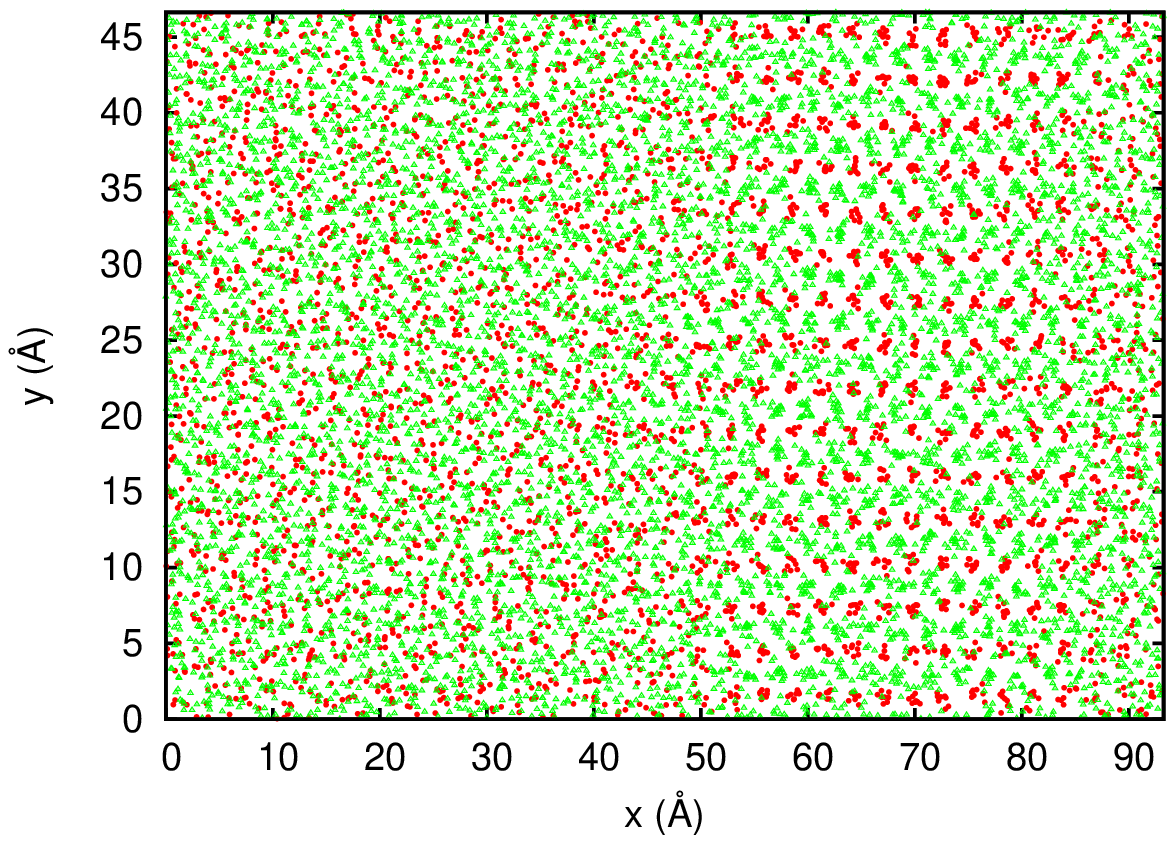}}%
        \caption{Determination of melting temperatures   
                based on two-phase coexistence molecular dynamics simulations. 
                \emph{Top}: Number of particles histogram represented as a function 
                of position along the direction parallel to the initial solid-liquid
                boundary.
                \emph{Bottom}: Atomic positions projected over the $x-y$ plane of
                 the simulation box. Calcium and fluor ions are represented 
                 with red and green dots, respectively.}
\label{fig6}
\end{figure}

Following previous works~[\onlinecite{cazorla07a,cazorla07b,cazorla09,cazorla11,cazorla12}],
we performed comprehensive $( N, V, E )$ two-phase coexistence MD simulations in order to 
determine the melting curve of cubic and orthorhombic CaF$_{2}$ under pressure. 
Starting with a supercell containing the perfect crystal structure (i.e. either fluoride
or PbCl$_{2}$-type), we thermalize it at a temperature slightly below the expected melting 
temperature for about $10$~ps. The system remains in a superionic state. 
The simulation is then halted and the positions of the atoms 
in one half of the supercell are held fixed while the other half is heated up 
to a very high temperature (typically five times the expected melting temperature) 
for about $60$~ps, so that it melts completely. With the fixed atoms still fixed, the 
molten part is rethermalized to the expected melting temperature (for about $10$~ps).
Finally, the fixed atoms are released, thermal velocities are assigned, and the whole system 
is allowed to evolve freely at constant $( N, V, E )$ for a long time 
(normally more than $100$~ps), so that the solid and liquid come into equilibrium. 
The system is monitored by calculating the average number of particles in slices of the cell 
taken parallel to the boundary between the solid and liquid. With this protocol, there is a 
certain amount of trial and error to find the overall volume which yields the coexisting solid 
and liquid system. (An example of a successful coexistence run is shown in Fig.~\ref{fig6}.)
Our simulations were done on cells containing $12,288$ atoms with the long axis being 
perpendicular to the initial liquid-solid boundary. 

In both cubic and orthorhombic structures, we calculated the
melting transition temperature $T_{m}$ at six different volumes. We find 
that our results can be very well fitted to the so-called Simon 
equation~[\onlinecite{simon29}]
\begin{equation}
T_{m}(P) = a \left( 1 + \frac{P}{b} \right)^c~, 
\label{eq:simon}
\end{equation}
where $a = 2044.05$~K, $b = 1.2049$~GPa and $c = 0.1202$ are the optimal values 
for the cubic phase, and $a = 389.19$~K, $b = 0.0180$~GPa and $c = 0.3159$ for the 
orthorhombic. Our predicted melting temperature at 
$P = 0$ is $2044$~($100$)~K, which turns out to be slightly larger than the experimental 
value $1690$~($20$)~K~[\onlinecite{mclaughlan67,mitchell72}]. At this temperature the 
calculated bulk modulus of the superionic phase is $\sim 57$~GPa, which is significantly 
smaller than the obtained for the perfect cubic crystal (see Table~II).    
Using the Simon equation, we estimate the corresponding zero-pressure melting 
slope to be $dT_{m} / dP = 203.97$~K~GPa$^{-1}$. We note that this quantity 
is considerably reduced under pressure, in constrast to the constant 
$dT_{s} / dP$ case reported in the previous section; at $P = 5$~GPa, for instance, we find  
$dT_{m} / dP \sim 40$~K~GPa$^{-1}$. 
Interestingly, the value of the calculated high-P melting slopes are
larger than those measured in alkaline-earth (AE) fluorides (e.g. LiF
and NaF)~[\onlinecite{boehler97}], in spite of the structural similarities between both 
type of structures. AE-fluorides (rock-salt) and CaF$_{2}$ (fluorite) have both an
fcc array of anions and differ only in the positions of the cations. A
possible cause for the larger melting slope of CaF$_{2}$ could be that
AE-fluorides melt directly from an ordered crystalline structure, while
CaF$_{2}$ melts from the superionic phase. Due to diffusion of F atoms, the
degree of atomic disorder is larger in superionic CaF$_{2}$ than in rock-salt
AE-fluorides. Therefore, it is smaller the corresponding entropy of
fusion (and hence so $\Delta H$). Consequently, according to the
Clasius-Clayperon relation, the slope of the melting line should be
larger in CaF$_{2}$.
Regarding the melting properties of CaF$_{2}$ in the orthorhombic phase, 
this exhibits a very low zero-pressure melting point of $\sim 400$~K (a temperature at 
which the corresponding crystal structure is metastable) and a overall steep 
$dT_{m} / dP$ slope (see Fig.~\ref{fig3}). 
It is worth noticing that the melting line of the orthorhombic phase 
intersects that of the cubic phase at the \emph{special} triple point
$(7.18~{\rm GPa}, 2580~{\rm K})$. 
At the triple point an increase occurs in the melting slope. This
phenomenon is most likely due to the volume change associated with the
solid-solid transition~[\onlinecite{errandonea02}].
Unfortunately, we do not know of any 
experimental data to compare with these results. 

Concerning previous $T_{m}$ estimations, we are just aware 
of works done by Zijiang~[\onlinecite{zijiang05}], Wang~[\onlinecite{wang10}], 
and Zeng \emph{et al.}~[\onlinecite{zeng08}] who employed similar pair-potential 
models than us here. The $P = 0$ melting temperatures 
predicted by those authors range from $1650$ to $2100$~K, which are 
in reasonable good agreement with our results. Nevertheless, we must note that 
the computational approaches used in those studies mainly consist of one-phase 
MD simulations and heuristic overheating arguments, both of which are well-known to 
produce very imprecise $T_{m}$ 
results~[\onlinecite{cazorla07a,cazorla07b,cazorla09,cazorla11,cazorla12,haskins12,hernandez10}].    
In fact, Zeng \emph{et al.}~[\onlinecite{zeng08}] find a melting temperature
of $990-1073$~K for the $Pnma$ phase of CaF$_{2}$ at $P = 10$~GPa, a result that 
appreciably differs from ours $T_{m} = 2867$~K obtained under identical 
conditions. The main reasons for such discrepancies may probably lie 
on the different methodologies employed and the complete omission of 
superionic effects in Zeng's work.

\subsection{Special triple points and other two-phase boundaries}
\label{subsec:boundaries}

\begin{table*}
\begin{center}
\label{tab:boundaries} 
\begin{tabular}{c c c c c c} 
\hline
\hline
$ $ & $ $ & $ $ & $ $ & $ $ & $ $ \\
$ $ & $ \frac{dT}{dP}~({\rm K/GPa})$ & $P_{c}~({\rm GPa})$ & $T_{c}~({\rm K})$ & $\Delta V~({\rm \AA^{3}/atom})$ & $\Delta H~({\rm eV/atom})$ \\
$ $ & $ $ & $ $ & $ $ & $ $ & $ $ \\
\hline
$ $ & $ $ & $ $ & $ $ & $ $ & $ $ \\
${\rm Fluoride-PbCl_{2}}$                       & $-315~(40)$ & $6.78~(5) $ & $1600~(10)$ & $-0.97~(2)$ & $0.031~(5)$ \\
$ $ & $ $ & $ $ & $ $ & $ $ & $ $ \\
${\rm Superionic~Fluoride-PbCl_{2}}$            & $650~(100)$ & $7.18~(5) $ & $1860~(10)$ & $-1.01~(2) $ & $-0.023~(5)$ \\
$ $ & $ $ & $ $ & $ $ & $ $ & $ $ \\
${\rm Superionic~Fluoride-Superionic~PbCl_{2}}$ & $ \infty  $ & $7.18~(5) $ & $2580~(10)$ & $-0.36~(2) $ & $0.00~(5)$ \\
$ $ & $ $ & $ $ & $ $ & $ $ & $ $ \\
\hline
\hline
\end{tabular} 
\end{center}
\caption{Thermodynamic data describing the unravelled solid-solid, solid-superionic, 
and superionic-superionic phase boundaries. 
The slope of the $P-T$ boundaries and corresponding changes in 
volume ($V$) and enthalpy ($H$) at the three \emph{special} triple points $(P_{c},T_{c})$ are 
reported.} 
\end{table*}

In previous sections, we have characterized four out of the seven two-phase boundaries 
shown in Fig.~\ref{fig3}. As it has been explained at the beginning of this section, 
we determined the remainder of boundaries (i.e. solid-solid, solid-superionic and 
superionic-superionic) based on fundamental thermodynamic considerations, 
one-phase MD simulations, and assuming linear pressure dependence  
in all of them. The results so obtained are summarized in Table~IV. 
As one can observe, the slope of the superionic fluoride-superionic
PbCl$_{2}$ boundary is infinite because the calculated 
enthalpy difference between the two superionic phases at 
$(7.18~{\rm GPa}, 2580~{\rm K})$ is zero (see Eq.~\ref{eq:clausius}). 
In the other two cases, we find that the slope of the 
superionic fluoride-PbCl$_{2}$ boundary is positive 
(i.e. $\Delta H < 0$ and $\Delta V < 0$) and roughly 
a factor of two larger in absolute value than of the fluoride-PbCl$_{2}$ 
boundary. In this last case, the resulting $dT / dP$ slope is negative 
because the differences in enthalpy and volume between the two crystals are of 
opposite sign (i.e. $\Delta H > 0$ and $\Delta V < 0$). In fact, \emph{a priori} 
one would expect the entropy of the fluoride~$\to$~PbCl$_{2}$ transformation  
to be positive, and hence so $\Delta H$, because higher symmetry structures 
(i.e. cubic CaF$_{2}$) in general imply lower entropy.  

Interestingly, we identify the presence of three \emph{special} triple points, 
$(P_{c},T_{c})$, in the thermodynamic region $6 \le P \le 8$~GPa and 
$1500 \le T \le 2750$~K (see Table~IV and Fig.~\ref{fig3}). 
These \emph{special} thermodynamic states are located at the intersections between 
three different phase boundaries, and the pressures and temperatures at which 
are predicted to occur in principle can be accessed in experiments. 
To this regard, information contained in Fig.~\ref{fig3} and Table~IV 
must be considered as highly valuable since identification of coexisting superionic 
and liquid phases turns out to be very challenging in practice. 
In fact, we just know of a couple of recent experimental works  
wherein coexistence between superionic and liquid phases has been suggested 
to happen in water upon very extreme $P-T$ conditions~[\onlinecite{goncharov09,sugimura12}].    

If one lifted the linear pressure dependence assumption from the solid-solid, 
solid-superionic and superionic-superionic phase boundaries, the location of 
the three \emph{special} triple points $(P_{c},T_{c})$ quoted in Table~IV 
will probably change. In order to quantify the magnitude of those variations 
and to assess so the accuracy in our results, one could for instance perform 
calculations of the Gibbs-Duhem integration type and exactly determine  
the involved multi-phase boundaries~[\onlinecite{kofke93,sanz04,mcbride12}]. 
Gibbs-Duhem integration calculations and other equivalent exact 
schemes~[\onlinecite{cazorla12,alfe02,haskins12}] however are computationally very 
intensive so that we opted for a more straightforward test. 
In particular, we computed the value of the $dT / dP$ slope at states 
$(P'_{c},T'_{c})$ found at halfway of the approximated linear multi-phase 
boundaries (see Fig.~\ref{fig3}), and checked whether these differed
appreciably or not from those reported in Table~IV. 
For the superionic cubic-superionic orthorhombic phase boundary, we find 
that the enthalpy difference between the two phases at 
$(7.18~\rm{GPa} , 2220~\rm{K})$ is zero implying also an infinite slope; 
assuming linear pressure dependence, therefore, seems to be adequate in this case. 
By contrast, at point $(6.98~\rm{GPa} , 1728~\rm{K})$ belonging to the superionic 
fluoride-solid PbCl$_{2}$ boundary we obtain a $dT / dP$ value that is roughly
twofold larger than the obtained at the corresponding \emph{special} 
triple point $(7.18~\rm{GPa} , 1860~\rm{K})$~. However, the extend of this 
last two-phase boundary is so reduced that we may still assume that the
resulting $(P_{c},T_{c})$ inaccuracies are reasonably small. 
Certainly, at state $(8.87~\rm{GPa}, 800~\rm{K})$ of the cubic-orthorhombic
phase boundary we find $dT / dP = -300~(30)$~K/GPa which is in very good agreement 
with the constant assumed value of $-315~(40)$~K/GPa reported in Table~IV. 
These outcomes come to show that assuming linear pressure dependence in 
all two-phase boundaries involving superionic and crystal structures 
provides very consistent results, in spite of the small errors introduced in 
the superionic cubic-solid orthorhombic boundary. Consequently, 
our \emph{special} triple point estimations can be safely considered as accurate.

\section{The role of repulsive and dispersion interactions on melting}
\label{sec:discussion}

In contrast to \emph{ab initio} electronic band structure methods, empirical 
and semi-empirical force fields may suffer from  
versatility and transferability issues. This means that an interaction 
model which correctly describes a set of properties may  
fail at reproducing others and/or the same under different thermodynamic
constraints. Actually, a considerable number of CaF$_{2}$ pairwise potentials 
are found in the literature each having been designed for a
distinct purpose~[\onlinecite{gillan90,wilson96,dick58,boulfelfel06,speziale02}]. 
Trying to derive the phase diagram for all of them would indeed be a tedious 
and extremely boring task. Fortunately, once the phase stability  
properties of a given interatomic potential are known it is possible  
to deduce those for other similar interaction models in a computationally 
efficient and physically insightful way. We refer here to the original free-energy 
perturbative approach developed by Gillan and collaborators and which has been 
successfully applied to the study of transition metals under extreme $P-T$ 
conditions~[\onlinecite{cazorla07a,cazorla07b,cazorla09,cazorla12,alfe02}].
In this section, we use Gillan's ideas to compute the shift in melting 
temperature caused by mild variations of the potential parameters employed 
through this work (i.e. under the general transformation $\lbrace X_{ij} \to X'_{ij} \rbrace$,
where $\lbrace X_{ij} \rbrace$ correspond to the set of parameters reported in Table~I). 
The motivation for this analysis is not only to gain access to the  
phase diagram features of other similar BMH potentials but to understand also 
the general role of short- and long-ranged interactions in melting. A brief 
description of Gillan's free-energy perturbative approach is provided next.  

For a given $P$ and $T$, the difference $G_{n}^{l s} \equiv G_{n}^l - G_{n}^s$ 
between the Gibbs free energies of the \emph{new} liquid and solid
(i.e. obtained with the \emph{new} set of potential parameters $\lbrace X'_{ij} \rbrace$) 
deviates from the corresponding difference $G_{0}^{l s} \equiv G_{0}^l - G_{0}^s$ of the 
initial reference liquid and solid 
(i.e. obtained with the reference set of potential parameters $\lbrace X_{ij} \rbrace$ shown in Table~I), 
and we write:
\begin{equation}
\label{eq:gibbsdiff}
G^{ls}_{n}(P,T)=G^{ls}_{0}(P,T)+ \Delta G^{ls}(P,T) \; .
\end{equation}
The shift $\Delta G^{l s} ( P , T )$ caused by changing the total-energy function from
$U_{0}$ to $U_{n}$, induces a shift in the corresponding melting temperature $T_{m} (P)$.
To first order, the latter shift is~[\onlinecite{alfe02}]:
\begin{equation}
\label{eq:correctiont}
\Delta T_{m} =\frac{\Delta G^{ls}\left(T_{m}^{0}\right)}{S^{ls}_{0}} \; ,
\end{equation}
where $S_{0}^{l s}$ is the difference between the entropies of the liquid and solid (i.e. the
entropy of fusion) of the initial reference system, and $\Delta G^{l s}$ is 
evaluated at the melting temperature of the initial reference system. 
The shift $\Delta G^{l s}$ is the difference of shifts of Gibbs free energies of the liquid and solid
caused by the shift $\Delta U \equiv U_{n} - U_{0}$.
Under constant volume and temperature, the shift of Helmholtz free energy $\Delta F$ arising from 
$\Delta U$ is given by the well-known expansion:
\begin{equation}
\label{eq:helmholtz_exp}
\Delta F = \langle \Delta U\rangle_{0} - 
\frac{1}{2} \beta \langle \delta\Delta U^{2}\rangle_{0} + \cdots \quad , 
\end{equation}
where $\beta\equiv 1/k_{\rm B}T$, $\delta\Delta U\equiv \Delta U - \langle \Delta U\rangle_{0}$, 
and the averages are taken
in the initial reference ensemble. From $\Delta F$, we obtain the shift of Gibbs free energy 
at constant pressure as:
\begin{equation}
\label{eq:isocorisobar}
\Delta G = \Delta F - \frac{1}{2}V\kappa_{T}(\Delta P)^{2} \quad ,
\end{equation}
where $\kappa_{T}$ stands for the isothermal compressibility and 
$\Delta P$ is the change of pressure caused by the replacement 
$U_{0} \to U_{n}$ at constant $V$ and $T$.

\begin{figure}
\centerline
        {\includegraphics[width=1.0\linewidth]{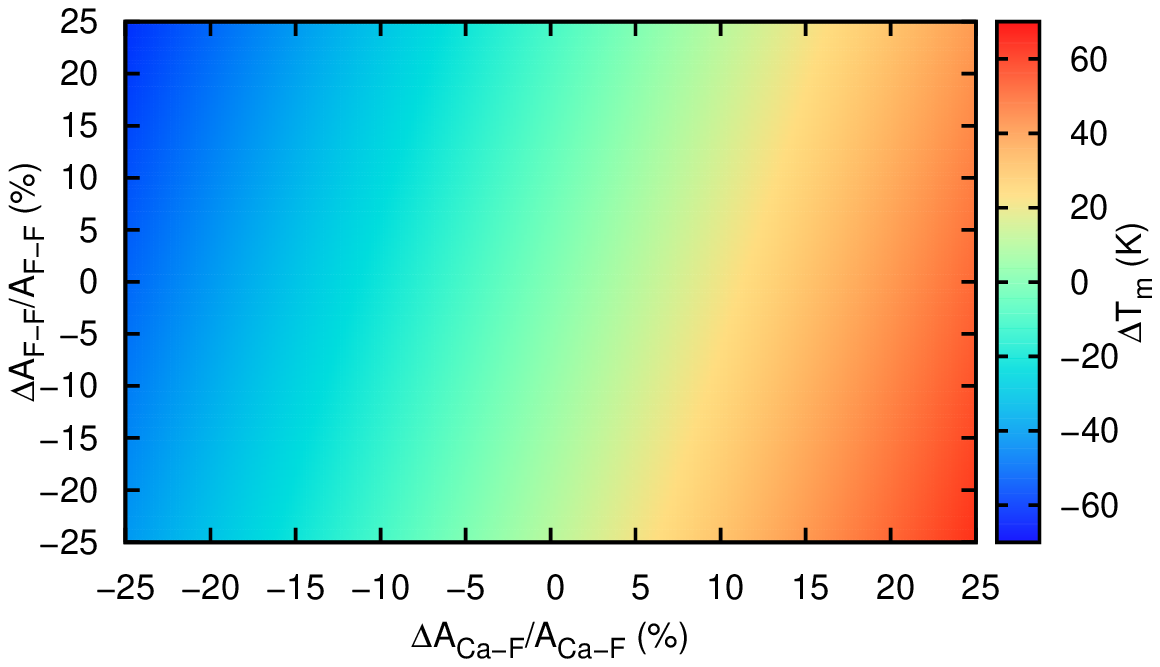}}%
        {\includegraphics[width=1.0\linewidth]{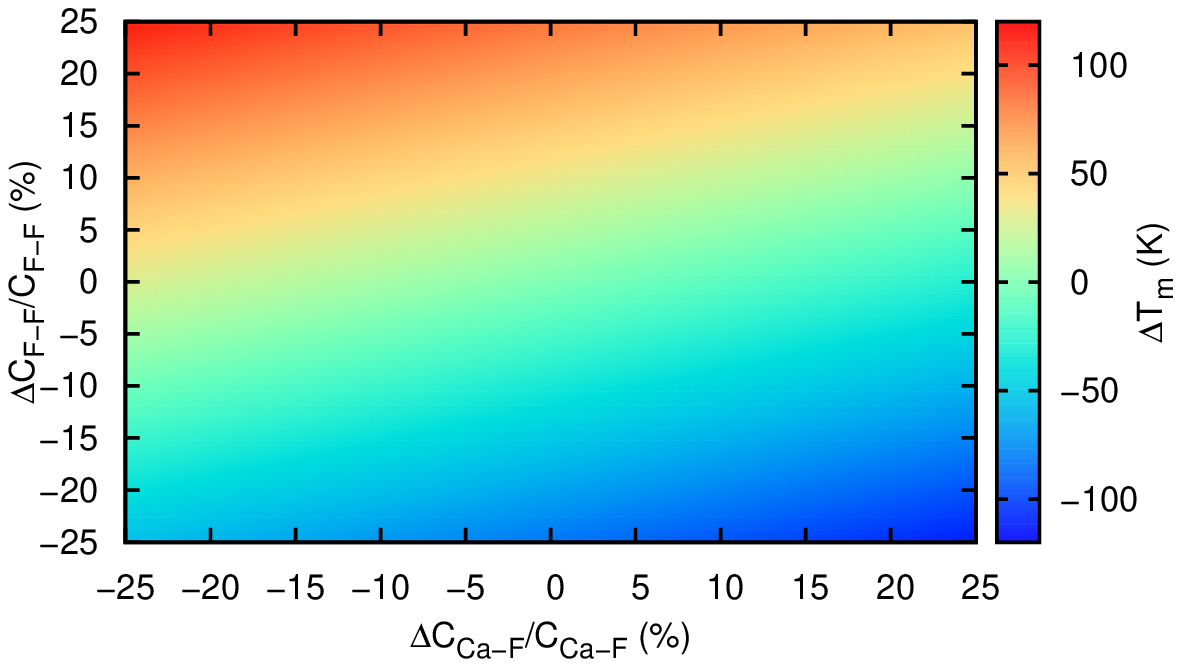}}%
        \caption{Shift of the melting transition temperature $\Delta T_{m}$ caused by  
                the variation of potential parameters at $P_{m} = 1.0$~GPa and 
                $T_{m} = 2180$~K. 
                \emph{Top}: Short-range repulsive $A_{\rm Ca-F}$ and
                $A_{\rm F-F}$ parameters are varied and the rest of parameters kept fixed.     
                \emph{Bottom}: Long-range attractive $C_{\rm Ca-F}$ and
                $C_{\rm F-F}$ parameters are varied and the rest of parameters kept fixed.   
                 }
\label{fig7}
\end{figure}

Our $\Delta T_{m}$ calculations were done over series of 
$\Delta X_{ij}/X_{ij}$ points ($\Delta X_{ij} \equiv X'_{ij} - X_{ij}$,
where $X_{ij}$ refer to the reference potential parameters reported in Table~I) 
generated in the range $-25~\% \le \Delta X_{ij}/X_{ij} \le 25~\%$ and
taken at 1~\% intervals (see Fig.~\ref{fig7}). The averages involved in these calculations
were computed over $500$ liquid and solid configurations generated 
in long one-phase MD simulations performed with the reference BMH potential. 
These MD simulations were carried out at the arbitrarily selected  
state $(1.0~{\rm GPa}, 2180~{\rm K})$ lying over the superionic fluoride-liquid 
phase boundary.  

Fig.~\ref{fig7} shows our $\Delta T_{m}$ results expressed as a function of
$\Delta A_{ij}/A_{ij}$ and $\Delta C_{ij}/C_{ij}$ variations (i.e. we have 
neglected $\rho_{ij}$ fluctuations). 
As one may appreciate, significant variations of the short-ranged 
and long-ranged parts of the interatomic potential in general have only a 
moderate effect on $T_{m}$. For instance, $A_{ij}$ and $C_{ij}$ relative variations of 
$\sim 25$~\% provoke at most melting temperature shifts of $\sim 100$~K. This 
result is surprising but at the same time reassuring in the sense that it adds robustness 
to our $T_{m}$ conclusions drawn in the previous section.  
Now, let us focus on the $\Delta T_{m}$ shifts caused by individual 
$A_{\rm FF}$, $A_{\rm CaF}$, $C_{\rm FF}$, and $C_{\rm CaF}$ 
variations. We note that due to the perturbative character of our approach 
the total melting temperature shift provoked by a general $\lbrace X_{ij} \to  X'_{ij} \rbrace$ 
transformation is equal to the sum of $\lbrace \Delta T_{m} \rbrace$ shifts caused 
by the individual $X_{ij}$ fluctuations. Concerning the short-ranged part of the 
interaction model (see top of Fig.~\ref{fig7}), we find that reduction of the 
$A_{\rm FF}$ parameter causes a positive melting temperature shift; by constrast, increase of
the same parameter leads to $\Delta T_{m} < 0$. This outcome comes to show that 
strengthening~(weakening) of the repulsive F-F interactions tends to further 
stabilize~(destabilize) the liquid phase. Interestingly, we observe 
the opposite trend for $A_{\rm CaF}$ : $\Delta A_{\rm CaF} < 0$ fluctuations lead to 
$\Delta T_{m} < 0$ (i.e. the liquid phase is energetically favored) whereas 
$\Delta A_{\rm CaF} > 0$ lead to $\Delta T_{m} > 0$ (i.e. the liquid phase is energetically 
disfavored). 
Upon a same $\Delta A_{ij} / A_{ij}$ change, we observe that the melting temperature 
shift obtained in the $A_{\rm FF}$ case is smallest (in absolute value)
(e.g. $\Delta T_{m} = -10$ and $51$~K for $\Delta A_{\rm FF} / A_{{\rm FF}}$
and $\Delta A_{\rm CaF} / A_{{\rm CaF}} = 25$~\%, respectively)
thus repulsive Ca-F interactions play a more dominant role in melting.   
Regarding the long-ranged attractive part of the interaction model 
(see bottom of Fig.~\ref{fig7}), the situation is the opposite
than just explained. In particular, $\Delta C_{\rm FF} < 0~(\Delta C_{\rm FF} > 0)$ 
changes lead to $\Delta T_{m} < 0~(\Delta T_{m} > 0)$ and 
$\Delta C_{\rm CaF} < 0~(\Delta C_{\rm CaF} > 0)$ to $\Delta T_{m} > 0~(\Delta T_{m} < 0)$. 
Moreover, upon a same $\Delta C_{ij} / C_{ij}$ variation the melting temperature shift 
obtained in the $C_{\rm FF}$ case is largest (in absolute value)
(e.g. $\Delta T_{m} = 83$ and $-29$~K for $\Delta C_{\rm FF} / C_{{\rm FF}}$
and $\Delta C_{\rm CaF} / C_{{\rm CaF}} = 25$~\%, respectively)
thus attractive F-F interactions play a more important role in melting.
As a summary of these results, we can state that  
short-ranged repulsive Ca-F and long-ranged attractive F-F contributions 
to melting are most notorious and that reduction~(increase) of the involved 
paremeters $A_{\rm CaF}$ and $C_{\rm FF}$ leads to further 
stabilization~(destabilization) of the liquid over the cubic superionic phase. 

Finally, we can use our results shown in Fig.~\ref{fig7} to predict, at least
at a qualitative level, the melting features corresponding to other
CaF$_{2}$ BMH potentials.  
In particular, we examine two parametrizations independently proposed by 
Gillan~[\onlinecite{gillan90}] and Boulfelfel~[\onlinecite{boulfelfel06}]. 
For the sake of simplicity, we consider here only
$\Delta T_{m}$ contributions stemming from the dominant 
$A_{\rm CaF}$ and $C_{\rm FF}$ parameters. 
For Gillan's parametrization, we obtain 
$\Delta A_{\rm CaF} / A_{{\rm CaF}} \sim -60$~\% and 
$\Delta C_{\rm FF} / C_{{\rm FF}} \sim 550$~\%. Actually, these values turn out to be 
exceedingly large so as to being treated within our perturbative approach 
however we can make a qualitative statement based on the size and 
sign of those deviations. In particular, the $C_{\rm FF}$ difference is eminently 
the largest and positive so that the resulting temperature correction is very 
likely to be positive and large (i.e. $T_{m} \gg 2180$~K).
In the case of Boulfelfel's potential, we obtain 
$\Delta A_{\rm CaF} / A_{{\rm CaF}} \sim 72$~\% and
$\Delta C_{\rm FF} / C_{{\rm FF}} \sim -100$~\%. These values are of opposite
sign to Gillan's and still too large so as to being analyzed 
with our method. Nevertheless, at the qualitative level, we can state that the 
resulting $\Delta T_{m}$ difference is very likely to be negative and small 
since the $C_{\rm FF}$ deviation is negative and only slightly superior 
than $A_{\rm CaF}$ (i.e. $T_{m} \lesssim 2180$~K). 

As a concluding remark to this section we want to mention that the
computational strategy just presented can also be applied to the analysis 
of multi-phase boundaries others than melting, and in general to the modelling 
of atomic interactions for derivation of phase diagrams.

\section{Conclusions}
\label{sec:conclusions}

To summarize, we have studied the phase diagram of CaF$_{2}$ under
pressure using classical atomistic simulations and a simple 
pairwise interatomic potential of the Born-Mayer-Huggings form.
Our results show that a rich variety of crystal, superionic and liquid phases 
coexist within the thermodynamic region $0 \le P \lesssim 20$~GPa 
and $0 \le T \lesssim 4000$~K. In particular, we find seven different 
two-phase boundaries for all of which we provide an accurate parametrization. 
Interestingly, three \emph{special} triple points are predicted to exist 
within the narrow and experimentally accessible thermodynamic range of 
$6 \le P \le 8$~GPa and $1500 \le T \le 2750$~K.
Indeed, we believe that these stimulating findings should encourage new 
experimental searches in CaF$_{2}$ under elevated $P-T$ conditions. 
Also, we have analyzed the role of short-ranged repulsive and 
long-ranged attractive atomic interactions in the prediction of 
melting points, with the finding that 
repulsive Ca-F and attractive F-F contributions are most notorious. 
In order to get rid of possible versatility and transferability
force-field issues, it would be very much desirable to conduct
\emph{ab initio} simulation studies similar 
to the one presented here. Work in this direction is already 
in progress within our group.

\acknowledgments
This work was supported by MICINN-Spain (Grants No. MAT2010-18113,
CSD2007-00041, MAT2010-21270-C04-01, CSD2007-00045 and FIS2008-03845) and computing 
time was kindly provided by CESGA.

\end{document}